\documentclass[a4paper,11pt]{article}
\usepackage{pos}

\title{Production asymmetry of $D$ and ${\bar D}$ mesons
in the LHCb fixed target experiment and intrinsic charm in the nucleon}

\ShortTitle{Production asymmetry of $D$ and ${\bar D}$ mesons and asymmetric intrinsic charm}

\author*[a]{Antoni Szczurek}
\author[a]{Rafa{\l} Maciu{\l}a}

\affiliation[a]{Institute of Nuclear
Physics, Polish Academy of Sciences, ul. Radzikowskiego 152, PL-31-342 Krak{\'o}w, Poland}

\emailAdd{antoni.szczurek@ifj.edu.pl}
\emailAdd{rafal.maciula@ifj.edu.pl}

\abstract{We discuss production of $D^0$ and ${\bar D}^0$ mesons in 
proton-nucleus collisions in fixed target experiments.
We include gluon-gluon fusion, intrinsic charm and perturbative
recombination. 
We compare rapidity and transverse momentum distributions
obtained within our approach with recent fixed target LHCb
experimental data.
All the mechanisms seem important for inclusive
production of $D^0 + {\bar D}^0$ mesons. The recombination mechanism 
seems crucial to understand asymmetry in production of $D^0$ and 
${\bar D}^0$ mesons.}

\FullConference{42nd International Conference on High Energy Physics (ICHEP2024)\\
18-24 July 2024\\
Prague, Czech Republic\\}

\begin{document}
\maketitle

\section{Introduction}

Recently the LHCb collaboration measured production of
$D^0$ and $\bar D^0$ mesons in proton-nucleus fixed target experiment
\cite{fixed_target_LHCb}. They observed an asymmetry in the production
of $D^0$ and $\bar D^0$. In general, there can be different reasons
of the asymmetry. One is asymmetric charm in the nucleon as due
to meson cloud. Another is a recombination mechanism \cite{BJM2002}.
Here we summarize our recent results for fixed target LHCb experiments
\cite{MS2022,GMS2024}.
We shall considered different mechanisms:
\begin{itemize}
\item gluon-gluon fusion, dominant in the collider mode.
\item mechanism of $c/{\bar c}$ knock-out, related to 
      inrinsic charm (IC) in the nucleon 
      as formulated in \cite{BHPS1980} or within meson cloud model.
\item recombination mechanism of perturbative nature.
\end{itemize}

\section{Sketch of the formalism}

\begin{figure}[!h]
\centering
\begin{minipage}{0.3\textwidth}
  \centerline{\includegraphics[width=1.0\textwidth]{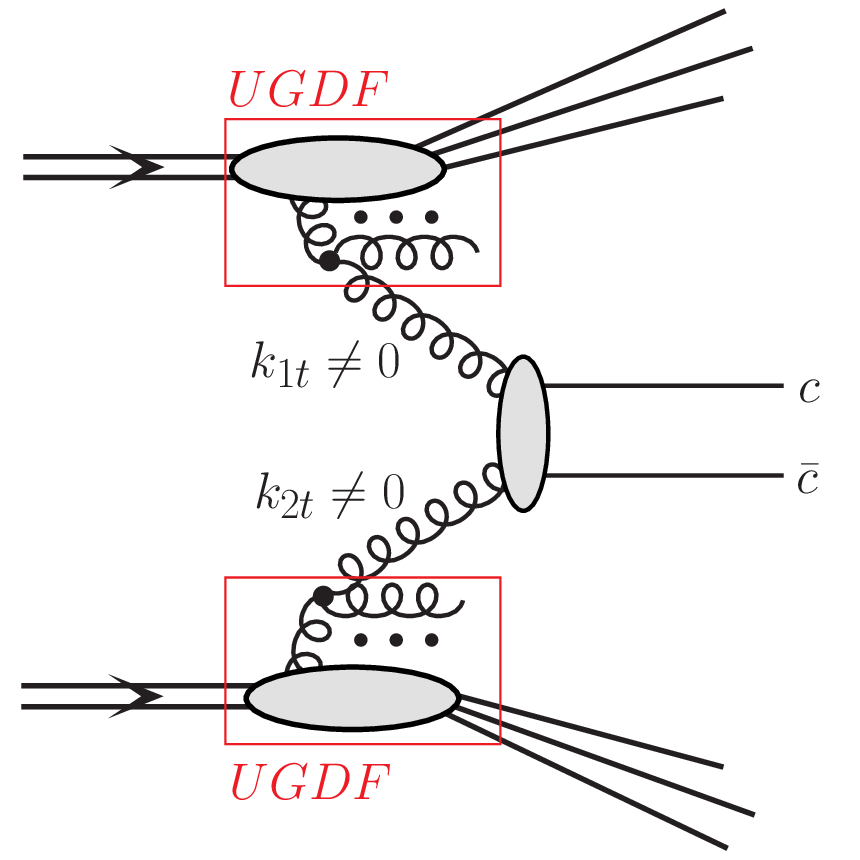}}
\end{minipage}
\caption{The dominant gluon-gluon fusion mechanism.}
\label{fig:ggfusion}
\end{figure}

In Fig.\ref{fig:ggfusion} we show the dominant at high 
energies mechanism of gluon-gluon fusion.

\begin{figure}[t]
\centering
\begin{minipage}{0.4\textwidth}
  \centerline{\includegraphics[width=1.0\textwidth]{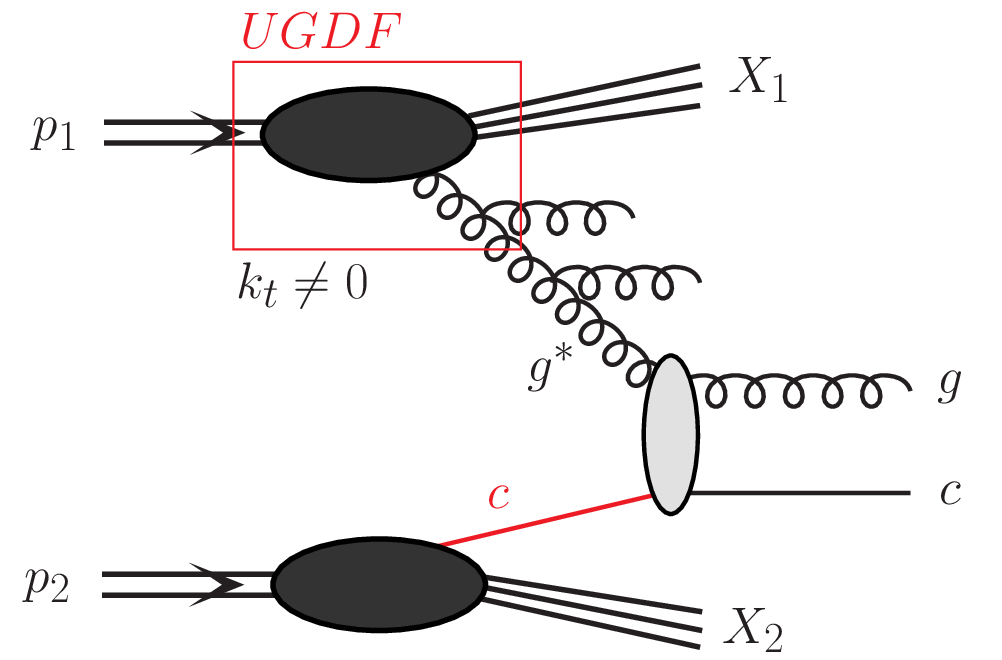}}
\end{minipage}
\begin{minipage}{0.4\textwidth}
  \centerline{\includegraphics[width=1.0\textwidth]{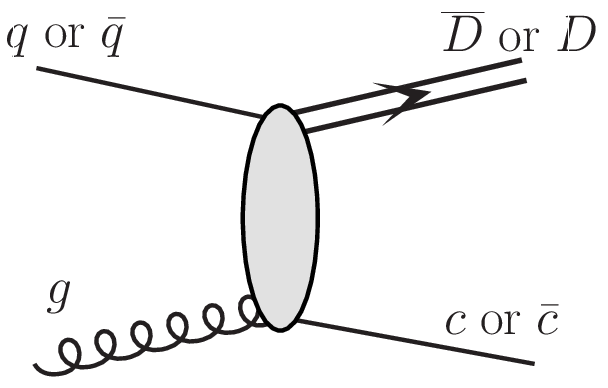}}
\end{minipage}
  \caption{
\small A diagrammatic representation of: the gluon - charm mechanism for charm production within the hybrid model, with the off-shell gluon and the on-shell charm quark in the initial state (left), and the generic leading-order diagrams for $D$ meson production via 
the BJM recombination model (right).
}
\label{fig:diagramIC}
\end{figure}

In addition to the dominant high-energy mechanism at ``our'' low 
energy $\sqrt{s}$ = 68.5 GeV we include also intrinsic charm 
contribution (see the left panel of Fig.\ref{fig:diagramIC}). 
Here the intrinsic charm quark or antiquark (identical distribution
in the most simple approach of Ref.\cite{BHPS1980}) 
is kicked off by the interaction with gluon from the proton. 
The second mechanism with the approximations made leads
to symmetric production of $c$ and $\bar c$ and in the consequence
identical production of $D^0$ and $\bar D^0$ mesons, also 
differentially.
In the presented analysis we included also perturbative recombination 
mechanism first proposed by Braaten, Jia and Mehen (BJM) 
\cite{BJM2002} (see the right panel of Fig.\ref{fig:diagramIC})).
As discussed in \cite{MS2022} this leads to $D^0$-${\bar D}^0$ asymmetry.

The gluon-gluon fusion is calculated in the $k_t$-factorization
mechanism as:

            \begin{eqnarray}
            \frac{d \sigma}{d y_1 d y_{2} d^2 p_{1,t} d^2 p_{2,t}} =
            \int \frac{d^2 k_{1,t}}{\pi} \frac{d^2 k_{2,t}}{\pi}
            \frac{1}{16 \pi^2 (x_1 x_2 s)^2} \; \overline{ | {\cal M}_{g^*g^* \rightarrow Q \bar Q} |^2} 
            \;\;\;\;\;\;\;\;\;\;\;\;\;\; \nonumber \\ 
            \;\;\;\;\;\;\;\;\;\;\;\;\;\;\;\;\;\;\;\;\;\;\;\;\;\;\;\;\;\;\;\;\;\;\;\;\; \times \;\;\;
            \delta^{2} \left( \vec{k}_{1,t} + \vec{k}_{2,t} 
                 - \vec{p}_{1,t} - \vec{p}_{2,t} \right) \;
            {\cal F}_g(x_1,k_{1,t}^2,\mu) \; {\cal
              F}_g(x_2,k_{2,t}^2,\mu) \; . \nonumber    
           \end{eqnarray}

Above ${\cal F}_g$ are unintegrated gluon distributions in proton or 
nucleus, respectively The same functional form is used for 
$^{20}N\!e$ as for the proton, i.e. no nuclear effects are taken into account
in order to concentrate on individual contributions (gluon-gluon fusion,
intrinsic charm, recombination).

The intrinsic charm contribution to $c$ and $\bar c$ distributions 
is calculated in a hybrid model as:

\begin{eqnarray}
d \sigma_{pp \rightarrow charm}(cg^* \rightarrow c g) = \int dx_1  \int \frac{dx_2}{x_2} \int d^2k_t \, \nonumber \\ \nonumber
\times \,\, c(x_1,\mu^2) \cdot {\cal{F}}_{g} (x_2, k_t^2, \mu^2) \cdot
d\hat{\sigma}_{cg^* \rightarrow  cg} \; ,
\end{eqnarray}
%
where $c(x_1,\mu^2)$ is collinear $c$-quark distribution.
Similar formula can be written for production of ${\bar c}$ antiquark.

The differential cross section for the recombination reads:

\begin{eqnarray}
\frac{d\sigma}{d y_1 d y_2 d^2 p_{t}} &=&
 \frac{1}{16 \pi^2 {\hat s}^2}
 [ x_1 q_1(x_1,\mu^2) \, x_2 g_2(x_2,\mu^2)
\overline{ | {\cal M}_{q g \to \bar{D} c}(s,t,u)|^2} \nonumber \\
&+& x_1 g_1(x_1,\mu^2) \, x_2 q_2(x_2,\mu^2)
\overline{ | {\cal M}_{g q \to \bar{D} c}(s,t,u)|^2} ]  \, . \nonumber
\label{cross_section}
\end{eqnarray}
In the formula above: $\overline{ | {\cal M}_{q g \to D c}(s,t,u)|^2} 
= \overline{ | {\cal M}_{q g \to ({\bar c} q)^n c} |^2} \cdot \rho$, where $\rho$ can be interpreted as a probability to form real meson. The parameter cannot be directly calculated and has to be extracted from the data.
Above $q$ is a collinear quark/antiquark distribution in proton or
neutron, components of the nucleus, with the flavour relevant for $D^0$ or ${\bar D}^0$ production and 
$g$ is a collinear gluon distribution.

The fragmentation from $c \to D^0$ or ${\bar c} \to {\bar D}^0$
is done using phenomenological fragmentation functions
known from $e^+ e^-$ collisions.
More details are explained in \cite{MS2022,GMS2024}.

\section{Selected results}

\begin{figure}[!h]
\begin{minipage}{0.5\textwidth}
  \centerline{\includegraphics[width=1.0\textwidth]{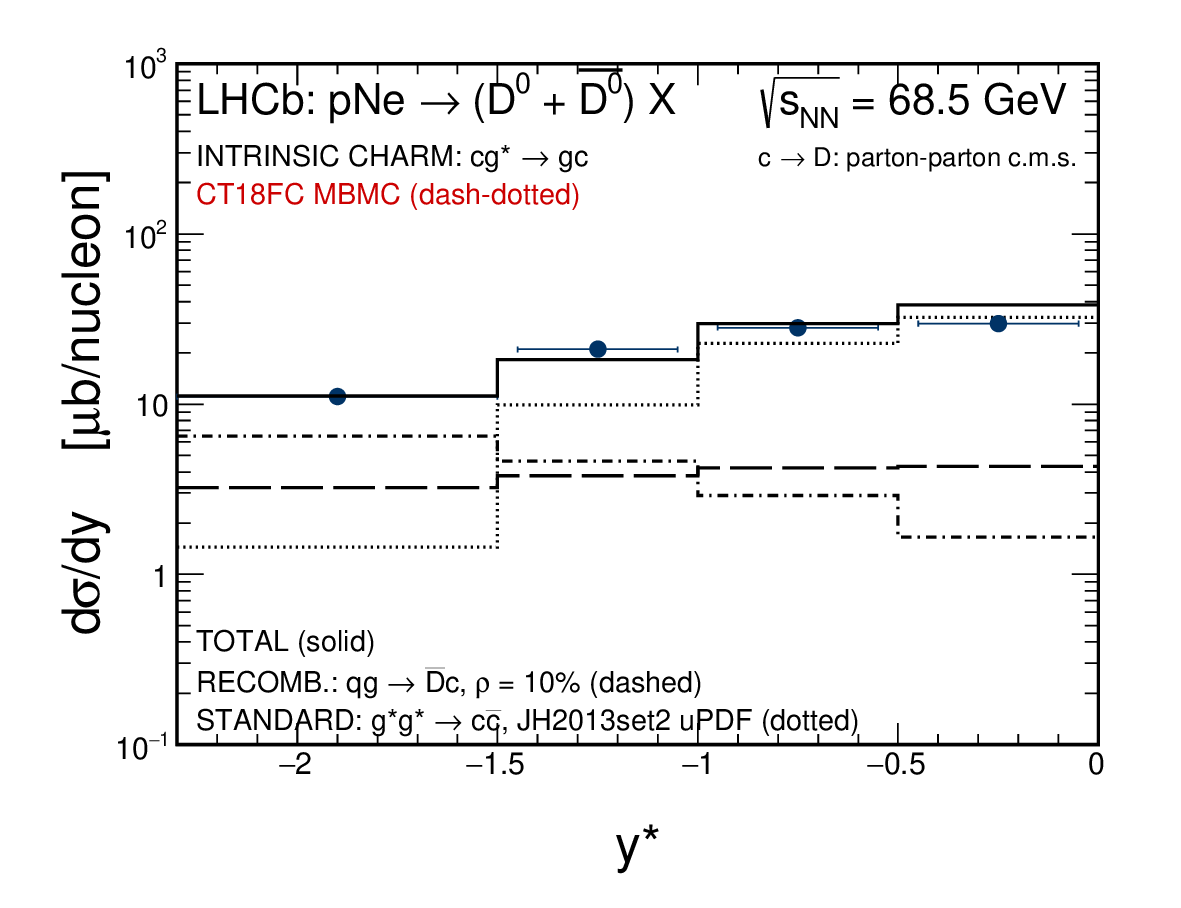}}
\end{minipage}
\begin{minipage}{0.5\textwidth}
  \centerline{\includegraphics[width=1.0\textwidth]{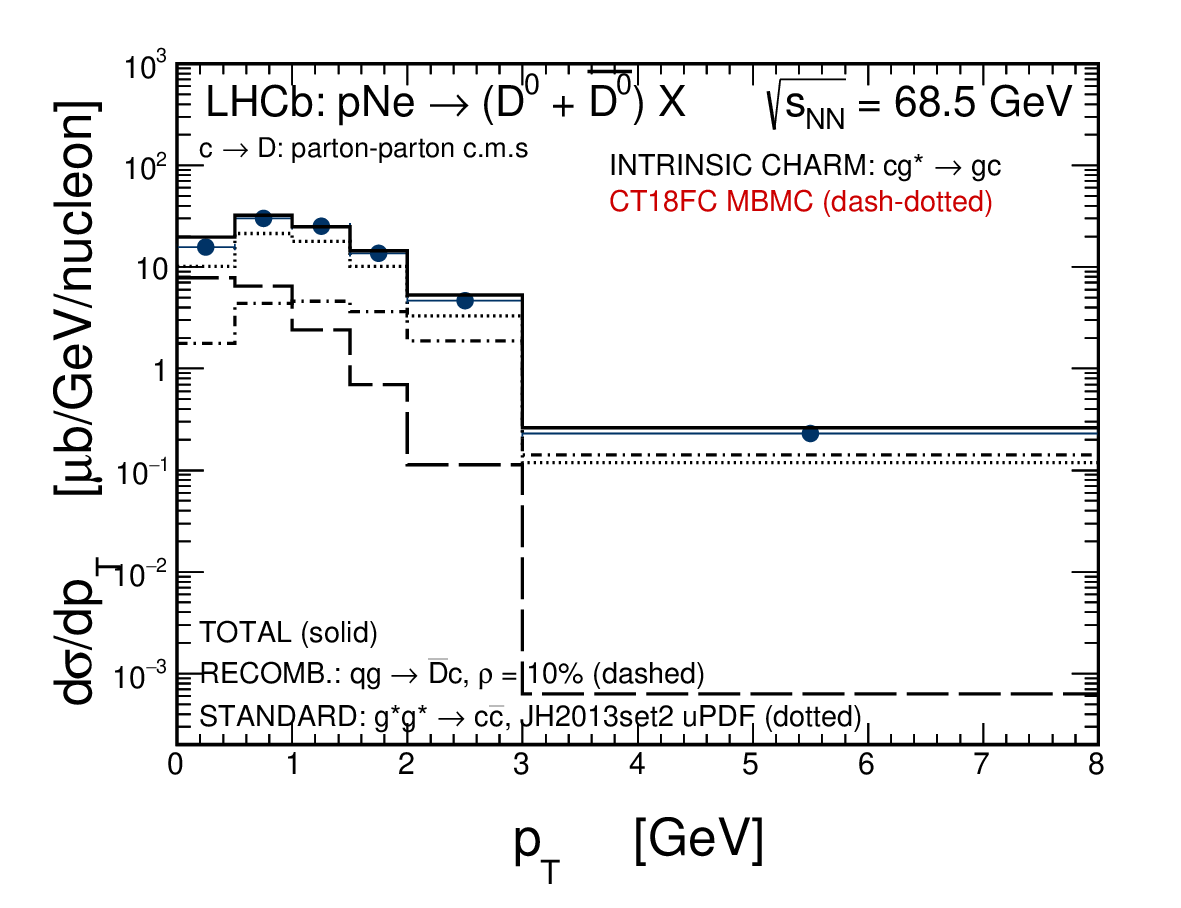}}
\end{minipage}\\
\begin{minipage}{0.5\textwidth}
  \centerline{\includegraphics[width=1.0\textwidth]{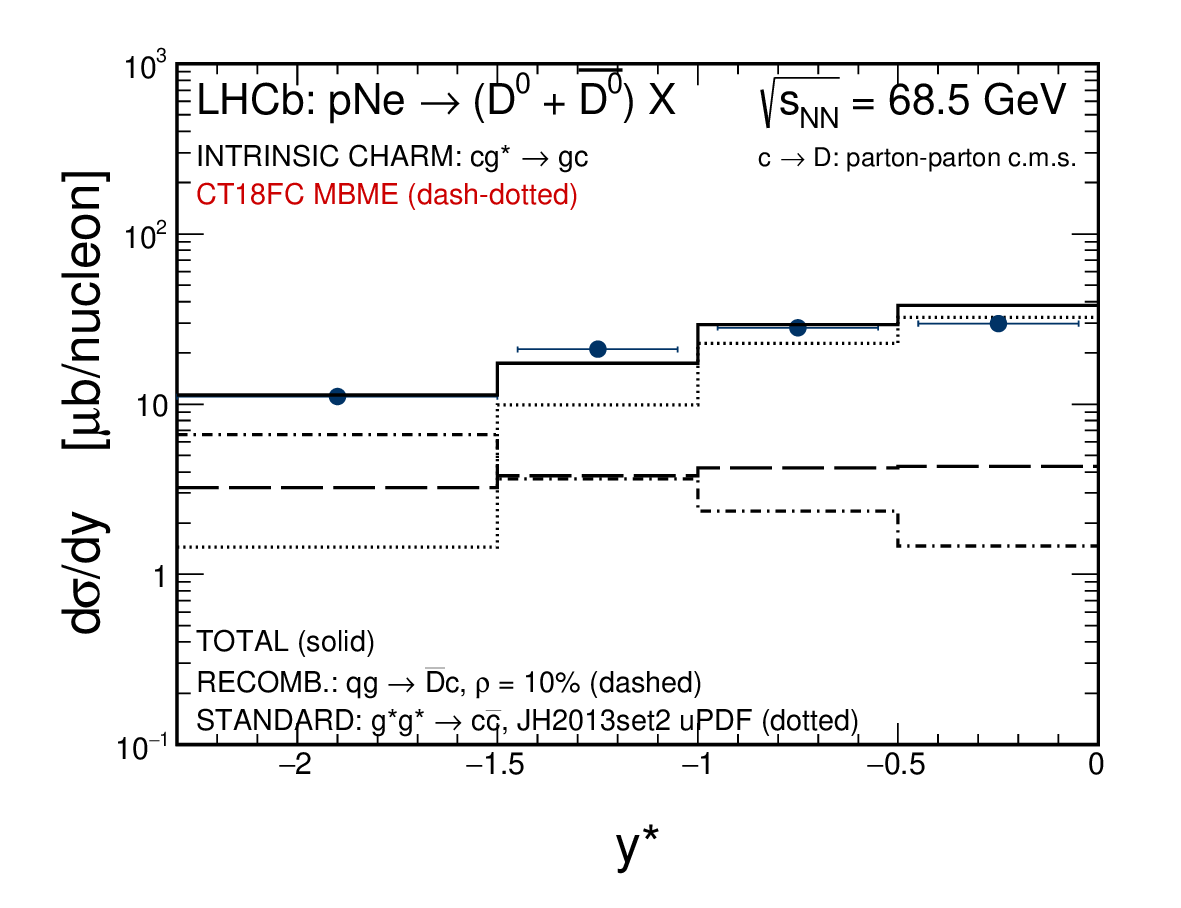}}
\end{minipage}
\begin{minipage}{0.5\textwidth}
  \centerline{\includegraphics[width=1.0\textwidth]{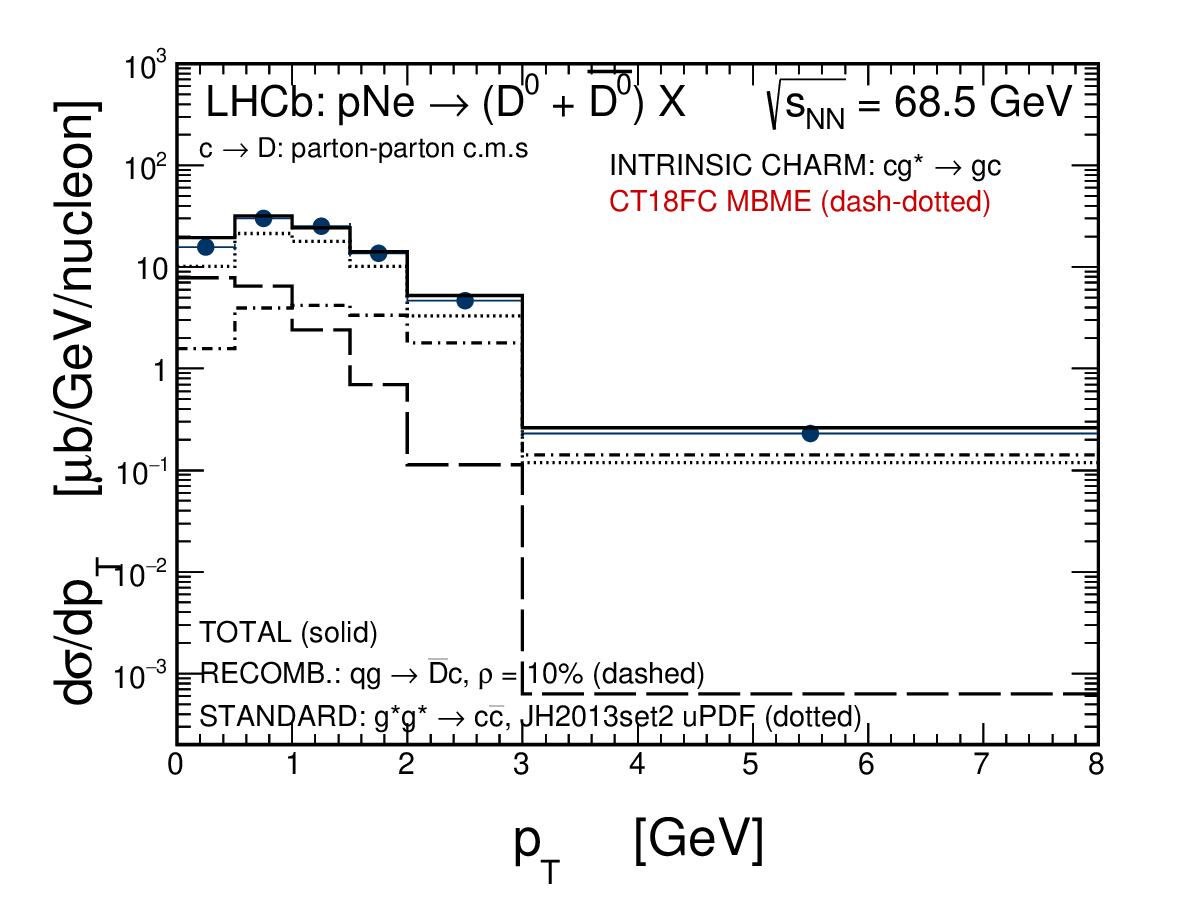}}
\end{minipage}
  \caption{\small The rapidity (left) and transverse momentum (right)
    distributions of $D^{0}$ meson (plus $\overline{D^{0}}$ antimeson)
    for $p+^{20}\!\mathrm{Ne}$ collisions at $\sqrt{s} = 68.5$ GeV
    together with the LHCb data \cite{fixed_target_LHCb}. The three
    different contributions to charm meson production are shown
    separately, including the standard $g^*g^*\to c\bar c$ mechanism
    (dotted), the gluon - charm contribution (dot-dashed) and the
    recombination component (dashed). The solid histograms correspond to
    the sum of all considered mechanisms. Results derived using the
    CT18FC MBMC (upper panels) and CT18FC MBME (lower panels)
    parametrizations are shown here.}
\label{fig:dist_difPDFs}
\end{figure}

In this section we present some representative results. In
Fig.~\ref{fig:dist_difPDFs} we present our predictions for the  rapidity
(left panel) and transverse momentum (right panel) distributions of
$D^{0}$ meson (plus $\overline{D^{0}}$ antimeson) in
$p+^{20}\!\mathrm{Ne}$ collisions at $\sqrt{s} = 68.5$ GeV together with
the LHCb data \cite{fixed_target_LHCb}. The inclusion of the gluon -
charm mechanism 
due to intrinsic charm improves the description of the experimental
data. In particular, in the region of the most backward rapidities both
the gluon - charm and the recombination contributions dominate over the
standard mechanism. There, the gluon - charm contribution is about
factor 2 and factor 5 larger than the recombination and the standard
mechanism, respectively. Considering the transverse momentum spectra, we
see that at low $p_{T}$'s the cross section is dominated by the standard contribution, however, for the highest transverse momentum bin the contribution from the gluon - charm mechanism starts to play important role. Although the data indicates that the inclusion of the IC contribution is important, we also found that the quality of the description of the LHCb data for the distinct models of asymmetric intrinsic charm is very similar. There are no big differences in the obtained normalization and shapes of the differential cross-sections for the inclusive single charm meson spectra between the CT18FC MBMC and CT18FC MBME parametrizations.

\begin{figure}[!h]
\begin{minipage}{0.5\textwidth}
  \centerline{\includegraphics[width=1.0\textwidth]{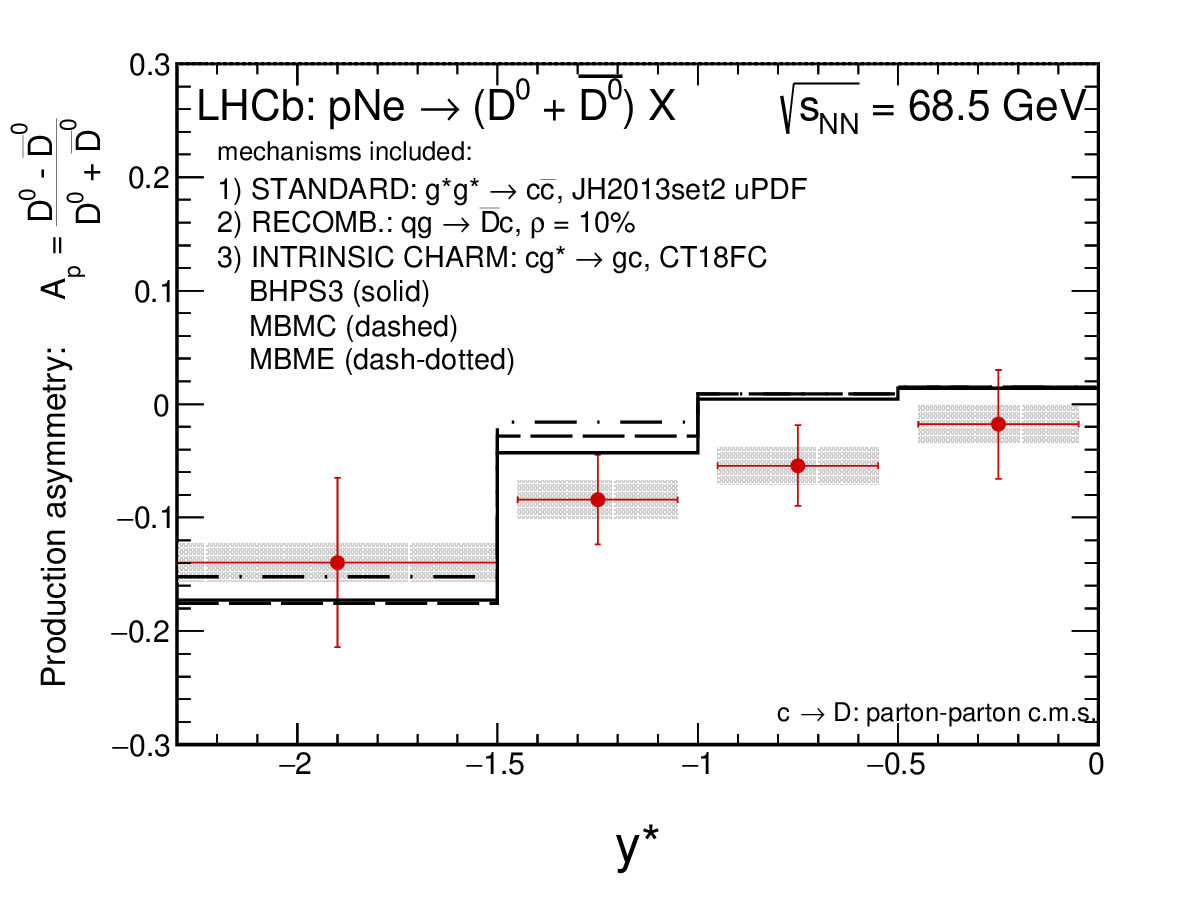}}
\end{minipage}
\begin{minipage}{0.5\textwidth}
  \centerline{\includegraphics[width=1.0\textwidth]{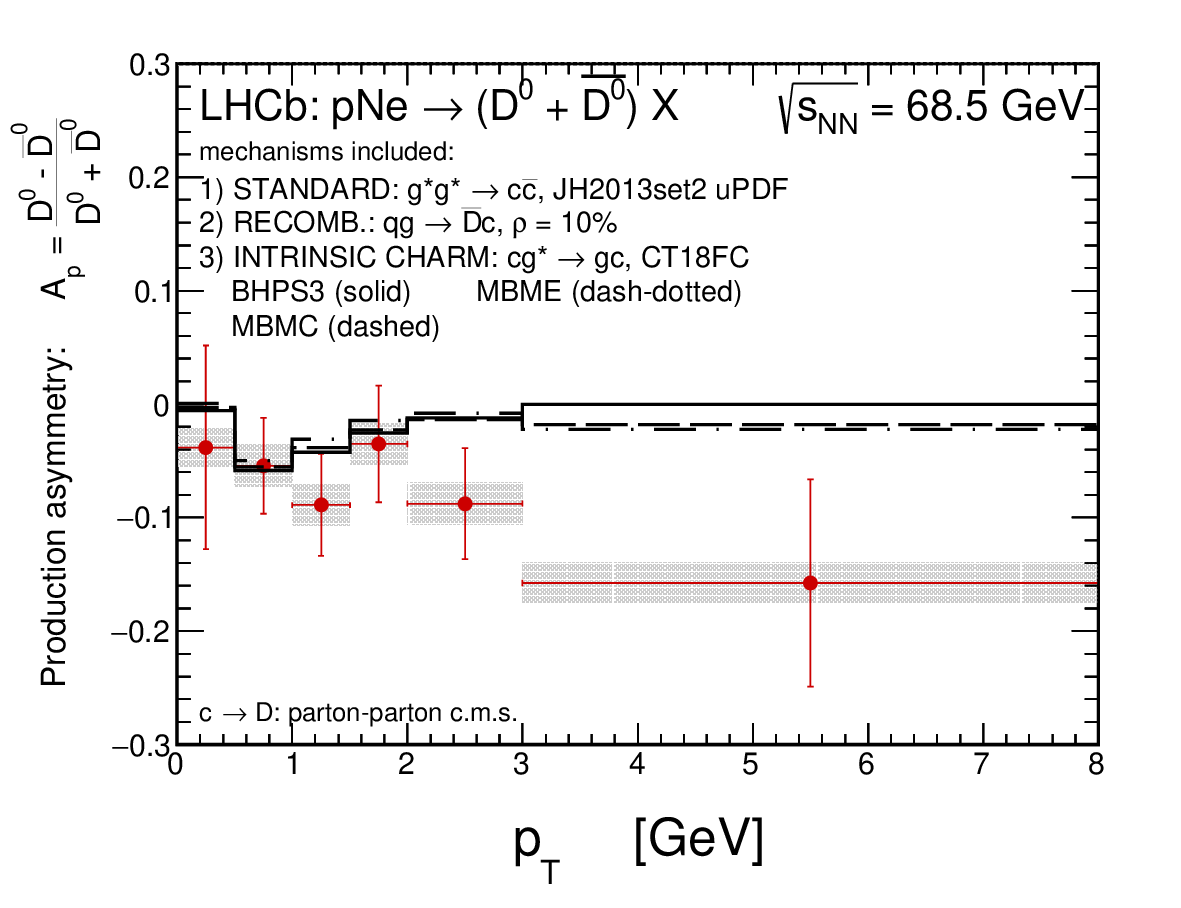}}
\end{minipage}
  \caption{
\small The production asymmetry $A_{p}$ for $D^{0}$-meson and
$\overline{D}^{0}$-antimeson as a function of rapidity (left) and
transverse momentum (right) for $p+^{20}\!\mathrm{Ne}$ collisions at
$\sqrt{s} = 68.5$ GeV together with the LHCb data
\cite{fixed_target_LHCb}. Results derived assuming the standard
$g^*g^*\to c\bar c$, gluon - charm and recombination mechanisms, and
different parametrizations for the intrinsic charm distributions are shown.
}
\label{fig:asym_sum}
\end{figure}

As was mentioned in the Introduction, the LHCb experimental data
indicates a sizeable production asymmetry of $D^0$ and $\bar D^0$ for
negative rapidities and large transverse momenta. Description of these
results in the full kinematical range is still a theoretical
challenge. The associated results of our predictions are presented in
Fig.~\ref{fig:asym_sum} for the distinct models of intrinsic charm
PDFs. The data can be described when both the recombination and
intrinsic charm mechanisms are taken into account. However, the
description of the data for large transverse momentum is still a
challenge. Clearly, more data in this kinematical region is needed in
order to constrain the IC and recombination mechanisms better.

\section{Conclusions}

At low $\sqrt{s}$ = 68.5 GeV p+A scattering all mechanisms: gluon-gluon fusion,
intrinsic charm and perturbative recombination mechanisms may play 
important role in the measured region. Each of the components is 
a bit unsure. We do not know very well unintegrated gluon
distribution relevant for the low energies (larger longitudinal
fractions). 
The intrinsic charm component depends e.g. on the probability of 
the $u u d c \bar c$ (proton) or $d d u c \bar c$ (neutron) Fock
components in proton or neutron, respectively.
While the 1 \% intrinsic charm knock-out mechanism improves 
the description of the LHCb rapidity distribution, the recombination
mechanism is necessary to generate the $D^0$-${\bar D}^0$
asymmetry as a function of $D$-meson rapidity. 
The $\rho \approx$ 0.1 allows to describe the asymmetry for the fixed 
target LHCb rapidity distribution. It is more difficult to describe
asymmetry as a function of $D$-meson transverse momentum.
This may be due to usage of leading-order colinear formalism which
may be not adequate for description of transverse momentum distribution.
It simply generates too small transverse momenta.
This could be easily improved by including primordial (nonperturbative)
distributions of partons (gluons, quarks, antiquarks) in the nucleon.

In this presentation we studied the case of symmetric intrinsic
charm distribution as in our first paper \cite{MS2022}.
In Ref.\cite{GMS2024} we included also asymmetric intrinsic charm
as dictated by the meson cloud picture. However, such an asymmetric 
mesonic IC is not able to describe alone (without the recombination 
mechanism) the measured $D^0 - {\bar D}^0$ asymmetries.
Therefore the recombination seems necessary.

Another mechanism, not discussed in the presentation at ICHEP2024, 
was discussed recently in \cite{V23}. There the production of
$D^0$ and ${\bar D}^0$ is asymmetric as $D^0$ and ${\bar D}^0$ come from different 
terms in the nucleon Fock expansion.

\end{document}